# Guiding Robust Valley-dependent Edge States by Surface Acoustic Waves

Zhen Wang[1], Fu-Kang Liu[1], Si-Yuan Yu[1,3], Shi-Ling Yan[1], Ming-Hui Lu[1, 3, 4]*, Yun Jing[2] and Yan-Feng Chen[1, 3]

[1]*National Laboratory of Solid State Microstructures and Department of Materials Science and Engineering, Nanjing University, Nanjing, 210093, China*

[2]*Department of Mechanical and Aerospace Engineering, North Carolina State University, Raleigh, North Carolina 27695, USA*

[3]*Jiangsu Key Lab. of Artificial Functional Materials, Nanjing University, Nanjing, 210093, China*

[4]*Collaborative Innovation Center of Advanced Microstructures, Nanjing, 210093 China*

*E-mail: luminghui@nju.edu.cn

Recently, the concept of valley pseudospin, labeling quantum states of energy extrema in momentum space, has attracted enormous attention because of its potential as a new type of information carrier. Here, we present surface acoustic wave (SAW) waveguides, which utilize and transport valley pseudospins in two-dimensional SAW phononic crystals (PnCs). In addition to a direct visualization of the valley-dependent states excited from the corresponding chiral sources, the backscattering suppression of SAW valley-dependent edge states transport is observed in sharply curved interfaces. By means of band structure engineering, elastic wave energy in the SAW waveguides can be transported with remarkable robustness, which is very promising for new generations of integrated solid-state phononic circuits with great versatility.

Valley, an intriguing and significant concept in condensed matter physics, stems from the extensive research of two dimensional (2D) hexagonal crystals, such as graphene, bilayer graphene, and transition-metal dichalcogenides, in recent years.[1-13] When the inversion symmetry is broken in two-dimensional (2D) hexagonal lattices, the berry curvatures will acquire opposite signs at K and K' points due to time-reversal symmetry. Subsequently, the electrons in different valleys own opposite anomalous velocities and move towards the opposite directions. Additionally, slow valley relaxation and dephasing processes, compared to electron spin, can be accessed due to the reason that intervalley scattering is suppressed by large momentum separation between different valleys.[2,3,12,14]

This merit can effectively reduce backscattering caused by many kinds of defects and disorders. Accordingly, valley degree of freedom has been proven to be distinctive from charge and spin, leading to the important field of research of spintronics. Recently, this concept was extended to phononic/photonic crystals (PnCs/PtCs) exhibiting valley-dependent transport of electromagnetic waves,[15-17] acoustic waves, and even mechanical/elastic waves.[18-23]

The macroscopic controllability enables the PnC to be a tractable tool for exploring acoustic analogues of complex quantum physics requiring atomic-scale manipulations. In addition, the low speed of sound of surface acoustic waves (SAWs) yields a far shorter wavelength than electromagnetic waves at the same frequency, which facilitates the miniaturization and integration of devices. Hence, it is worthwhile to investigate the valley-dependent effect in SAW PnCs. Ultimately, the extraordinary valley-dependent edge states could be potentially utilized to manipulate directional propagation of SAW for real-world applications.

Although the SAW valley-dependent edge states promise an exciting application prospect in SAW integrated devices, this specific topic has been relatively unexplored. Relevant efforts are mainly focused on the valley states for airborne sound and Lamb waves.[18-23] In this study, valley states are introduced into SAW waveguides. The valley states are obtained via opening an original elastic Dirac degeneracy of a PnC by breaking the spatial inversion symmetry. Time-reversal symmetry is maintained as no effective gauge field is present, *e.g.*, an external electronic/magnetic field or a dynamic modulation.[24-25] The specific vortex valley states can be excited from the corresponding chiral sources in the body of SAW PnCs. Furthermore, valley-dependent edge states transport with backscattering being strongly suppressed is also achieved by means of two types of SAW PnCs arranged with mirror symmetry along the interface. Such a passive SAW waveguide offers a promising design scheme for integrated solid-state phononic circuits and advanced acoustic devices with high signal fidelity (backscattering suppression).[26-28]

We obtain a band gap by breaking the $C_{6v}$ symmetry of an SAW PnC in honeycomb lattice for the introduction of valley states.[29] Solid nickel (Ni) pillars are arranged as a honeycomb structure on the surface of Y-cut lithium niobate (LiNbO$_3$) as shown in Fig. 1(a). The constitutive parameters of Ni and LiNbO₃ in the calculation are shown in Table I. $a_1$ and $a_2$ are lattice vectors in a graphene lattice, where $a$ (50μm) is the lattice constant. Pillars A and B represent the two atoms of hexagonal lattices. Full-wave simulations are carried out by a commercial finite-element solver (COMSOL Multiphysics). Floquet periodicitiy is applied as the boundary conditions of the unit cells. When the diameters and heights of pillars A and B are identical ($d_A=d_B=0.4a$, $h_A=h_B=0.5a$), there exists degenerate points at K and K' as the grey lines shown in Fig. 1(b). Once breaking the inversion symmetry by varying $d_A$ and $d_B$, *i.e.* $d_B/d_A=0.8$, the degeneracy can be opened and a band gap is formed from about 32.6MHz to 34.4MHz. The lower band-gap edges are labeled as bands I and III, and the upper ones are labeled as bands II and IV in black lines as shown in Fig. 1(b). Accordingly, the valley states with the feature of chirality corresponding to bands I-IV can be achieved in SAW systems and the chiral y-displacement fields are illustrated in Fig. 2.

TABLE I. Parameters of Ni and LiNbO₃ used in numerical simulations.

|  | Ni | LiNbO₃ |
| --- | --- | --- |
| Density (kg/m³) | 8900 | 4700 |
| Young's modulus (GPa) | 219 | 203 |
| Poisson's ratio | 0.31 | 0.31 |

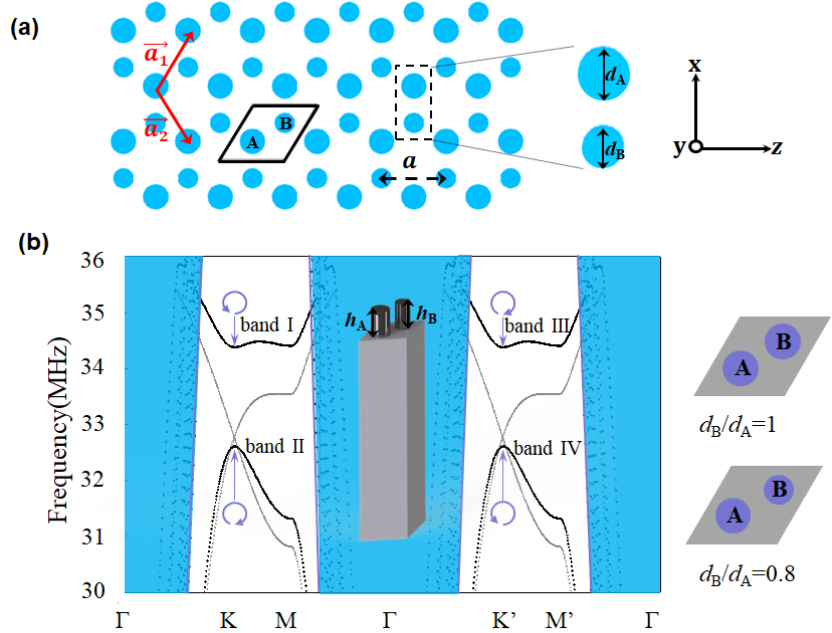

**FIG. 1.** Overlooking schematic view of the SAW graphene sheet and elastic dispersions around K and K'. (a) The solid Ni pillars are positioned in a graphene lattice with lattice vectors $a_1$ and $a_2$, where $a$ is the lattice constant. The $d_A$ and $d_B$ are the diameters of pillars A and B, respectively. (b) Gapless and gapped bulk dispersions for the cases $d_B/d_A$=1 (grey lines) and $d_B/d_A$=0.8 (black lines).

The chiral y-displacement fields for valley states can be clearly identified by the directions of energy flux vectors as shown in Fig. 2. If we only focus on the energy flux vectors at band I and band IV, it is evident that the energy flux vectors around the atoms A and B are along the same direction as the blue arrows shown in Figs. 2(a) and 2(d). Clearly, pillar B at band I and pillar A at band IV vibrate in a radially breathing manner which only has temporal out-of-plane displacement. In contrast, pillar A at band I and pillar B at band IV present peer chiral elastic energy flow rotating in the xz-plane in the time-domain. The similar chiral characteristics for valley states are shown in Figs. 2(b) and 2(c). When examining y-displacement fields of the two types of states at K point, elastic energy flow evolves in the opposite chirality as the blue arrows shown in Figs. 2(a) and 2(c). The opposite chiral features are also present as the blue arrows in Figs. 2(b) and 2(d). The existence of chiral elastic energy flow is an important characteristic of valley states, and is considered an analogue to pseudospin.[26,30,31] By utilizing these valley-induced pseudospins, researchers have achieved electromagnetic[15-17,32] and acoustic transmission line[19-22,33] with robust, valley-dependent transport of wave energy.

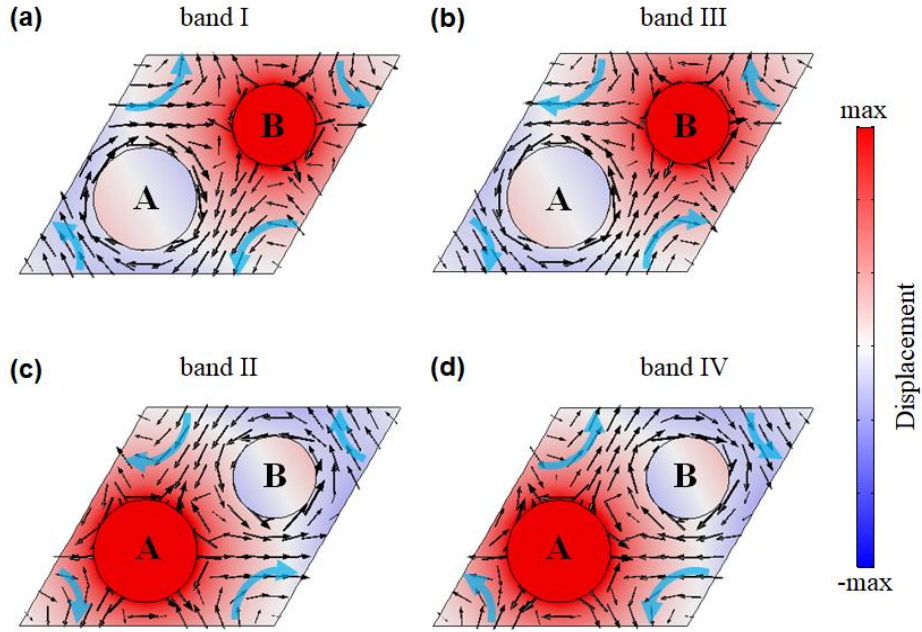

**FIG. 2.** Phase diagram locked with specific vortex features. (a) and (c) Phase profiles of the valley states at K. (b)

and (d) Phase profiles of the valley states at K'. Black and blue arrows show the directions of elastic energy flow.

The valley states with different elastic vortex fields can be selectively excited by appropriate chiral sources. A ring source is located at the center of a regular hexagon splitter as shown in the inset of Fig. 3(a), which allows us to study chiral transport with high fidelity even in the case of chiral excitation state. There are six output ports in this splitter (labelled as 1-6). In this case, the hexagonal SAW PnC is in accordance with geometry of the first Brillouin zone as shown in Fig. 3(b). The absorbing boundary condition is applied at the edges of the hexagon splitter. When a clockwise excitation at 32.6MHz is loaded on the ring source, the clockwise elastic energy can only travel along the ΓK direction and be allowed to flow out from the 1, 3, 5 output ports as shown in Fig. 3(c). In contrast, the anticlockwise elastic energy can only travel along the ΓK' direction and be allowed to flow out from the 2, 4, 6 output ports as shown in Fig. 3(d), if an anticlockwise excitation at 32.6MHz is loaded on the ring source. These observations from the regular hexagonal splitter model confirm the valley-dependent transportation with different chiral polarizations.

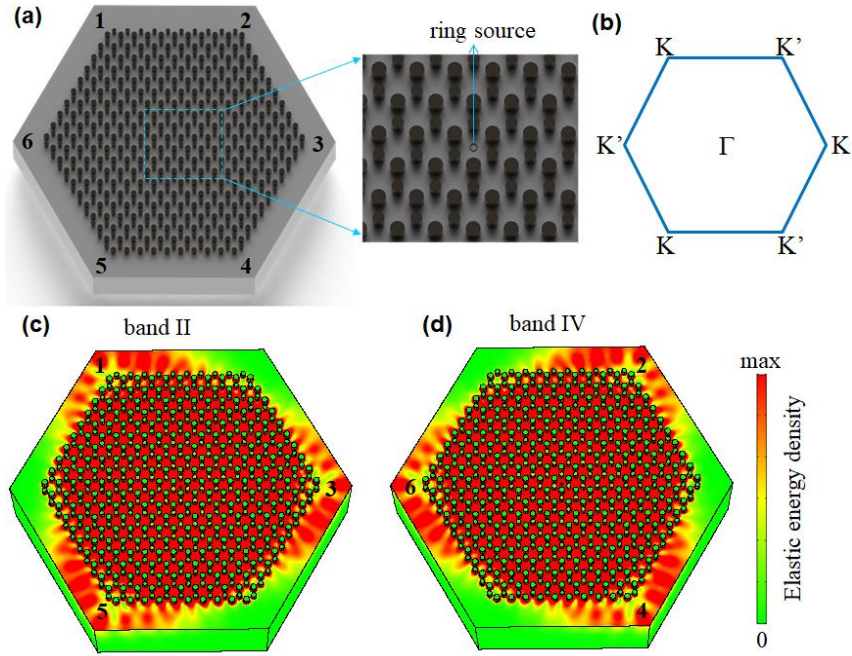

**FIG. 3.** Valley-selective excitation for the chiral valley states. (a) A schematic of regular hexagonal splitter with six output ports (labelled as 1-6). A ring source is located at the center of the splitter (see inset). (b) The first Brillouin zone of graphene sheet. (c) and (d) The elastic energy fields of two different chiral valley states in the hexagonal splitter.

The SAW valley-dependent edge states can be calculated in supercell models. In the simulation, Floquet periodicity along the z-axis and absorbing boundary conditions along the x-axis are applied at the sides of the supercell. The elastic dispersions of edge states along two kinds of zigzag-edged graphene boundaries (AB-BA and BA-AB) are shown in Figs. 4(a) and 4(b). The states $\phi_{\text{I,II}}^{\pm}$ and $\phi_{\text{II,I}}^{\pm}$ label those valley-dependent edge states[15-19] hosted by the interfaces I-II and II-I respectively, and they are displayed in Figs. 4(c) and 4(d). The y-displacement fields of $\phi_{\text{I,II}}^{-}$ and $\phi_{\text{I,II}}^{+}$ along the graphene boundary (AB-BA) are symmetric about the z axis as indicated in Fig. 4(c). Note that the edge states, *i.e.* $\phi_{\text{II,I}}^{+}$ and $\phi_{\text{II,I}}^{-}$, are in fact anti-symmetric about the z axis and they cannot be excited by a symmetric elastic source near the entrance of the waveguide. Therefore it is called a deaf state[34] as shown in Figs. 4(b) and 4(d). Here, we mainly focus on the SAW valley-dependent edge states, *i.e.* $\phi_{\text{I,II}}^{-}$ and $\phi_{\text{I,II}}^{+}$, as shown in Fig. 4(a). The gray dots in the band gap denote bulk states which are dissipated in the substrate. In the SAW system, although the edge states with positive group velocity ($+V_g$ at $+k$) and negative group velocity ($-V_g$ at $-k$) both exist in the band gap, only one of them can be excited independently if only one source is placed at one port of the one-dimensional waveguide. For example, if a source is placed at the right ports of the waveguides as illustrated in Figs. 5(a) and 5(b), only SAW valley-dependent edge state at $-k$ can be excited and it propagates leftwards, or vice versa.

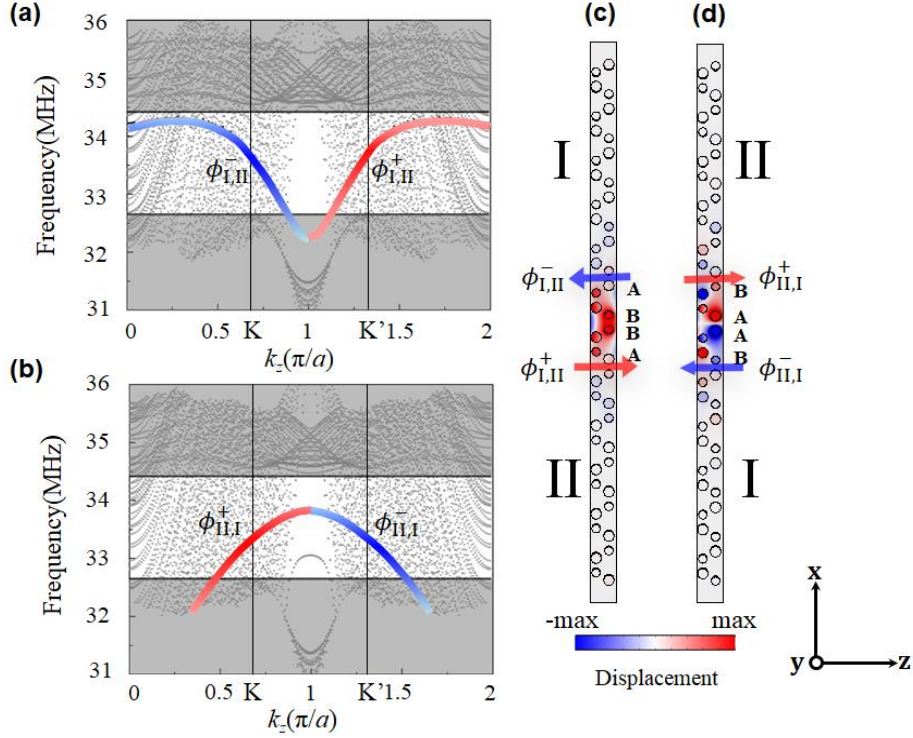

**FIG. 4.** Numerical validations of the valley-dependent edge states. (a) and (b) The states $\phi_{I,II}^{\pm}$ and $\phi_{II,I}^{\pm}$ label the valley-dependent edge states hosted by the interfaces I-II and II-I. (c) and (d) The symmetrical and antisymmetrical y-displacement fields of valley-dependent edge states hosted by the interfaces I-II and II-I. The domains I and II with $d_B/d_A$=0.8 are mirror-symmetrical with each other about z axis.

Based on above analysis, the existence of valley-dependent edge states in the band gap can be achieved through careful engineering of geometrical parameters of our SAW PnCs. In order to confirm the robust transport of these edge states in SAW systems, two different waveguides are constructed. Two models with a straight and a Z-shape waveguides are shown in Figs. 5(a) and 5(b). Corresponding fields of elastic energy are also presented, respectively. In the simulations, line sources are placed at the right entrance of the waveguides and the same monochromatic SAW excitation (33.2MHz) is applied. The absorbing boundary conditions are applied at the sides of the PnCs. Elastic energy of the valley-dependent edge states is seen strongly localized around the waveguides yielding almost loss-free transmission. The transmission spectra of these two waveguides are plotted in Fig. 5(c). The transmission through the valley-dependent Z-shape waveguide is nearly unaffected by the sharp corners. However, such a bending defect can cause significant disturbance on the elastic energy transportation in the bandgap-guiding PnCs waveguide,[18,23] which could include largely reduced transmission. Hence, the valley-dependent waveguide clearly distinguishes itself from the bandgap-guiding phononic crystal waveguide with the advantage of robust SAW transportation. It is worth noting that the transmission has some offset from the band gap. The reason is that the dispersion of valley-dependent edge states is not gapless and there is still a mini band gap near the upper edge of the original band gap.[35] The same phenomenon was observed in photonic crystals as well.[16]

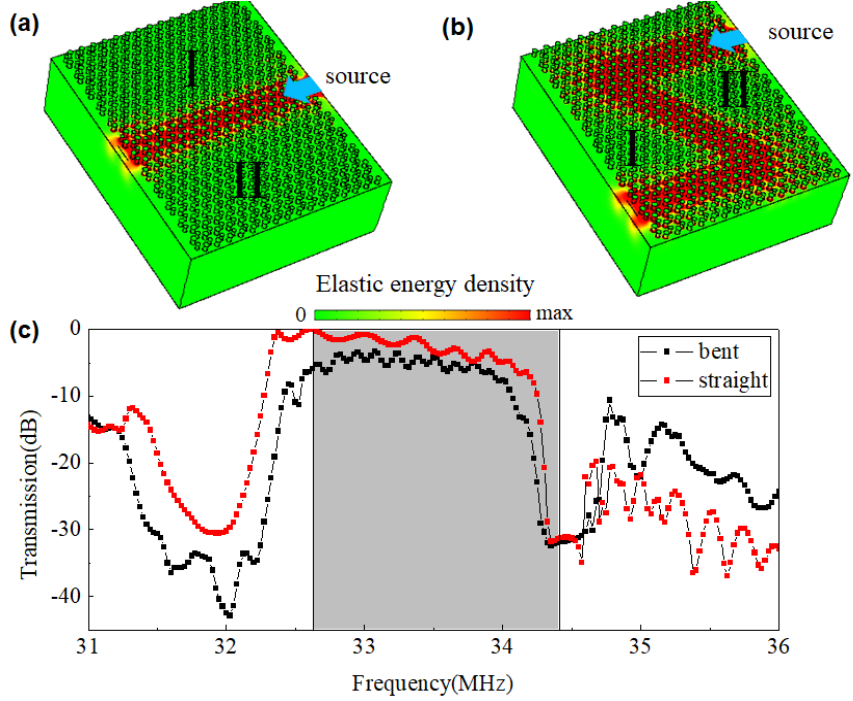

**FIG. 5.** Backscattering suppression for the SAW valley-dependent edge states from sharp corners. (a) and (b) The elastic energy fields along straight and bent interfaces simulated at 33.2MHz. (c) Transmission of the Z-shape path (black squares), compared with that for a straight channel sample (red squares).

In conclusion, we proposed and numerically investigated SAW valley-dependent edge states by breaking the inversion symmetry of honeycomb lattice. The SAW valley states at K or K' can be excited selectively by proper chiral sources. The valley-dependent edge states with backscattering suppression are demonstrated at bent interfaces. This robust transmission of elastic energy has the advantages of low loss and relatively wide bandwidth. The SAW valley-dependent properties revealed here could inspire further applications in *e.g.* on-chip acoustic sensing,[36] phonon-photon interactions[37,38] and acoustic signal processing.[39-41]

**Acknowledgments** This work is jointly supported by the National Key R&D Program of China (Grant No. 2017YFA0303702 and No. 2017YFA0305100,), the National Natural Science Foundation of China (No.51702152, No.51732006, No.11625418, No.11474158, No.11804149, No.51472114 and Grant No.51721001). We also acknowledge the fund from the Priority Academic Program Development of Jiangsu Higher Education (PAPD).